# Possibility of a new order parameter driven by multipolar moment and Fermi surface evolution in CeGe

Karan Singh and K. Mukherjee*

School of Basic Sciences, Indian Institute of Technology Mandi, Mandi 175005, Himachal Pradesh, India

*E-mail: kaustav@iitmandi.ac.in

Polycrystalline CeGe is investigated by means of DC and AC susceptibility, non-linear DC susceptibility, electrical transport and heat capacity measurements in the low temperature regime. This compound shows two peaks at low magnetic field around $T^I \sim 10.7$ and $T^{II} \sim 7.3$ K due to antiferromagnetic ordering and subsequent spin rearrangement respectively. Investigation of non-linear DC susceptibility reveals a presence of higher order magnetization which results in the development of a new order parameter around $T^I$. This leads to a lowering of symmetry of the magnetic state. The order parameter increases with decreasing temperature and stabilizes around $T^{II}$. Consequently, the symmetry of the magnetic state is preserved below this transition. Heat capacity and resistivity results indicate the presence of a gap opening around $T^I$ on portion of Fermi surface, due to evolution of the Fermi surface. Magnetoresistance behavior and violation of Kohler's rule suggest that the evolution of Fermi surface changes the symmetry of magnetic state. The observation of new order parameter (which is of second order) is also confirmed from the Landau free energy theory.

**Introduction**

The physics of 4$f$ electrons and its coupling with conduction electrons have resulted in novel phases, like antiferromagnetic state, spin density wave, unconventional superconductivity, nematicity, heavy fermions, Fermi liquid behavior etc. in different compounds [1-4]. Understanding the underlying mechanism for the occurrence of such phase remains a challenge in the field of the strongly correlated electron systems. In this context, the Ce-based intermetallic compounds are extensively studied because the 4$f$ level of $Ce^{3+}$ ions lie close to the Fermi level. The coupling of the 4$f$ electrons via the conduction electrons results in a magnetic ground state through a mechanism known as the Ruderman-Kittel-Kasuya-Yosida (RKKY) exchange interaction. In contrast, hybridization between the localized moment and conducting electrons gives rise to a non-magnetic ground state and this phenomenon is known as Kondo effect. Under the application of external perturbations like magnetic field, pressure, doping etc.; the competition between RKKY interaction and Kondo effect results in the observation of unique properties [5-7]. In recent years, substantial investigations are being carried out in the area of second order phase transition near quantum critical point (QCP), which is obtained by suppression of magnetic state at zero temperature. Furthermore, investigations of the presence of higher order magnetization have received a great deal of attention in recent years. As per literature reports, in some systems, magnetic ordering lowers the symmetry of magnetic state. This results in higher order magnetization due to presence of multiple interactions, leading to the development of a new order parameter [8-9]. The lowering of symmetry of the magnetic state might cause some changes in Fermi surface leading to a partial gap opening at Fermi surface [10-11]. Also, the gap opening might be a consequence of the formation of new Brillouin zone boundaries [11-12]. The interplay between higher order magnetization and gap opening at Fermi surface might reveal some new and interesting physical phenomena and may warrant an attention for understanding such behavior. In this context, CeGe is an interesting 4$f$ electron based system though; there are only few literature reports about this compound. This indicates that the magnetic state of this compound is quite complex [13-17]. Recent studies on single crystals of CeGe report the presence of magnetic anisotropy as well as gap opening [16]. However, to the best of our knowledge, a detailed study on polycrystalline CeGe is still lacking in literature. Hence, it would be interesting to explore extensively the magnetic and electronic properties of a polycrystalline CeGe. Special emphasis is given on the low field magnetic measurements to discern the magnetic ground state; as large applied fields might mask the intrinsic signatures of an inhomogeneously magnetized system.

     In this work, we report the results of DC and AC susceptibility, non-linear DC susceptibility, heat capacity and electrical transport measurement on polycrystalline CeGe. Our results show that the compound exhibits two anomalies around 10.7 and 7.3 K at low field. The first anomaly arises due to antiferromagnetic ordering while the second one is due to the subsequent spin rearrangement after the ordering. Presence of higher order magnetization around 10.7 K signifies a change of symmetry of the magnetic state. This results in development of a new order parameter due to the presence of multiple spin interactions. Also, a gap opening is

noted around 10.7 K due to evolution of Fermi surface because of a change in symmetry of the magnetic state. This evolution of Fermi surface is also reflected in magnetoresistance behavior and Kohler's plot analysis. Observation of new order parameter (which is of second order) is reconfirmed from the Landau free energy theory.

**Result and discussion**

**X-ray diffraction**

Figure 1 shows the powder x-ray diffraction pattern of this compound at room temperature. The obtained pattern indicates that the compound is crystallographically single phase. The pattern is also analyzed by the Rietveld profile refinement using FullProf suite software. The compound crystallizes in orthorhombic structure (space group Pnma). The obtained lattice parameters are $a$ = 8.358 (7) Å, $b$ = 4.082 (3) Å and $c$ = 6.028 (4) Å. The obtained pattern and the parameters are in analogy to that reported in [16].

**DC, AC and non-linear DC susceptibility study**

Figure 2 (a) and (b) shows the temperature ($T$) dependent (in the range 1.8 – 20 K) DC magnetic susceptibility in the magnetic field ($H$) range 0.005 – 7 T under zero field cooling (ZFC) and field cooled warming (FCW) conditions. As noted from figure 2 (a), two clear peaks are observed around $T^I$ ~ 10.7 K and $T^{II}$ ~ 7.3 K, at 0.005 T in the ZFC curve. Additionally, a large irreversibility is also observed between ZFC and FCW curves at the same field. As the magnetic field is increased, the peak around $T^I$ does not shift in temperature up to 1 T, but peak around $T^{II}$ initially broadens and gets suppressed above 0.1 T. Above 1 T, a single peak around $T^I$ is noted, which is shifted to the lower temperature with increasing magnetic field. Also, the strong irreversibility observed at lower field is absent. The inverse magnetic susceptibility of the compound at 0.1 T is fitted with Curie Weiss law in temperature range 150-300 K (inset of figure 2 (a)). The effective magnetic moment ($\mu_{eff}$) and Curie Weiss temperature ($\theta_p$) obtained from fitting are ~ 2.4$\mu_B$ and~ -34 K respectively. The experimental value of $\mu_{eff}$ matches with the effective moment of free ion of $Ce^{3+}$ and reflects the presence of the localized moments. The negative value of $\theta_p$ indicates to the dominance of antiferromagnetic interactions. Thus, it can be said that the peak around $T^I$ arises due to the antiferromagnetic ordering which is also consistent with the observed shifting of $T^I$ down in temperature, above 1 T. The presence of antiferromagnetic ordering around $T^I$ is in accordance to literature reports [15-16], however, there are no reports about the second peak which is observed at low fields. In order to collect further information, the magnetic field response of isothermal magnetization (*M*) curves at different temperatures is measured (as shown in figure 2 (c)). A significant hysteresis is observed at temperatures below $T^{II}$ and at low fields (below 0.4 T). The magnitude of the hysteresis increases with decreasing temperature below $T^{II}$ (inset of figure 2 (c)). This observation suggests the existence of ferromagnetism which might arise due to the presence of possible short range correlations among magnetic moments. Inset of figure 2 (b) shows the magnetic field response of $\Delta M/H$ (where $\Delta M = M_{FCW} - M_{ZFC}$) at 1.8 K. As stated above, and also observed from this figure, the bifurcation is significant only below 1 T. The observed irreversibility in magnetization can arise due to anisotropy effect. For a single crystal of CeGe, it has been reported that the anisotropy effect is minimal in low field whereas, it increases at high fields [16]. In our case, for polycrystalline CeGe, the bifurcation is suppressed above 1 T due to the suppression of weak ferromagnetism, possibly, because of onset of long ranged antiferromagnetic ordering among magnetic moments.

To cross-check the features observed in DC susceptibility, temperature dependent AC susceptibility at different frequencies and in presence of different DC fields are measured. Figure 3 (a) shows the temperature response of the real part of AC susceptibility ($\chi'_{ac}$) measured in 4 ($10^{-4}$) T AC field along with superimposed DC fields of 0.005 and 1 T. In analogy to DC susceptibility, two peaks are also noted near $T^I$ and $T^{II}$. Under superimposed DC field of 1 T, the peak around $T^{II}$ is suppressed. These peaks are also noted in the imaginary part of AC susceptibility ($\chi''_{ac}$) due to dissipation of magnetic energy arising out of spin dynamics (shown in inset of figure 3 (a)). At 1 T, both the peaks are suppressed, indicating towards the development of collinear magnetic ordering. Additionally, it is noted that the peaks show insignificant shift in temperature with increase in frequency (inset of figure 3 (b)), thereby ruling out the freezing mechanism of spins [18]. Figure 3 (b) shows the magnetic field dependent AC susceptibility ($\chi'_{ac}$) at frequency 931 Hz and AC field of 1 ($10^{-4}$T). In paramagnetic region (above $T^I$), $\chi'_{ac}$ decreases with increasing magnetic field, in both direction. However, below $T^I$, the shape of the curve changes drastically. It is observed that with increasing magnetic field, $\chi'_{ac}$ decreases and attains minima ~ ±1 T. Above ±1 T, $\chi'_{ac}$ increases and attains maxima ~ ±5 T and again decreases above ±5 T. However, below $T^{II}$, the features of the curve changes drastically and $\chi'_{ac}$ increases continuously with increasing magnetic field. Hence, our observation gives an indication of the presence of complex behavior arising out of the competing magnetic interaction which undergoes a change as the temperature is decreased.

In order to investigate the possibility of the presence of higher order magnetization in this complex magnetic state, the non-linear DC susceptibility (as a probe of higher order spin correlations) is studied, using the protocol as given in Ref [19, 20]. Figure 3 (c) show the $H^2$ dependent $M/H$ curves at selected temperature upto 15 K. Below $T^I$, it is noted that $M/H$ decreases linearly with $H^2$ and at low field (below 1 T), the curves sharply increases with decreasing magnetic field due to the presence of the ferromagnetic component. Such upturns are also noted in other Ce based compounds like $CeRu_2Si_2$ and $Ce_{1-x}Y_xRu_2Si_2$ [21]. Such behaviors of curves suggest that there is presence of significant higher order magnetization in whole field range. For extracting the higher order magnetization coefficient, a good fitting of $H^2$ dependent $M/H$ curves (upper inset of figure 3 (c)) are obtained in field range 2-7 T with the equation as:

$$M/H = \chi_1 + \chi_3 H^2 + \chi_5 H^4 \ldots \quad (1)$$

where $\chi_1$ is linear susceptibility, $\chi_3$ and $\chi_5$ are non-linear susceptibility. The obtained $\chi_1$ from this fitting replicates to that observed from direct DC susceptibility measurement and obtained value of $\chi_5$ is very small (not shown). Lower inset figure 3 (c) shows the temperature response of $\chi_3$. It is seen that a sign change from negative to positive take place just below $T^I$, and $\chi_3$ increases with decreasing temperature. Around $T^{II}$ and below it, a finite $\chi_3$ is noted. The growth of $\chi_3$ indicates the presence of higher order magnetization associated with quadrapole moment [22-23]. The development of quadrapole moment across $T^I$ suggests that symmetry of magnetic state is lowered due to the presence of an excited quartet ground state along with doublet state [8-9]. This results in quadrapole-quadrapole coupling and develops an order parameter which increases with decreasing temperature. The observation of finite $\chi_3$ around $T^{II}$ is an outcome of dominating quadrapole-quadrapole interactions, as compared to dipolar antiferromagnetic interactions. Hence, it can be said that the observed ferromagnetic component might arise because of short ranged correlations among quadrapolar magnetic moments. At high field (above 1 T), antiferromagnetic dipolar interactions dominate resulting in the observation of long ranged magnetic ordering and suppression of the ferromagnetic component. Hence, it can be said that the low field magnetic behavior of this compound is quite complex due to the presence of higher order magnetization, which possibly develops a new order parameter.

**Heat capacity study: Evidence for gap opening**

Figure 4 (a) shows heat capacity divided by temperature ($C/T$) as a function of temperature, in the field range 0–14 T. A jump in heat capacity is observed around $T^I$ at 0 T. No peak is observed around $T^{II}$, but a deviation in the curve is noted around that temperature. As the magnetic field is increased, it is noticed that there is no shift in the peak around $T^I$. But, beyond 1 T, the peak around $T^I$ broadens and is shifted towards the lower temperature, which is analogy with the DC susceptibility. The observation of this feature is due to the commencement of long ranged antiferromagnetic dipole moment which competes with quadrapolar moment. The value of the change in heat capacity at 0 T is ~ 7.0 J/mol K, which is small as compared to mean field value of 12.5 J/mol-K for a free $Ce^{3+}$ ion. The reduction of value could be due to partial screening of $Ce^{3+}$ localized moment via conduction electrons. Magnetic contribution of heat capacity ($C_{mag}$) is calculated after subtracting electronic ($\gamma$) and phonon ($\beta$) contributions in paramagnetic region from total heat capacity, using the equation

$$C/T = \gamma + \beta T^2 \quad \ldots\ldots (2)$$

and extrapolating the fitting of equation to lowest temperature from 25 K (Inset Fig. 3 (a). From the nature of the heat capacity curve (below 25 K), it is noticed that electron and phonon contribution dominates over Schottky type excitation in the temperature range of our fitting. The obtained value of $\gamma$ is 0.267 ±0.002 J/mol-$K^2$, which signifies that this compound can be classified as an intermediate heavy fermionic system. Upper inset of figure 4 (b) shows the temperature dependent magnetic entropy ($S_{mag}$) at 0 T. The $S_{mag}$ is calculated from integration of $C_{mag}$. Around $T^I$, $S_{mag}$ is found to be ~ 3.13 J/mol-K, which is 0.54 times of theoretical value of 5.76 J/mol-K. With decreasing temperature, $S_{mag}$ decreases and its value around $T^{II}$ is ~ 0.37 J/mol-K. Also, as noted from the temperature response of heat capacity curve at 0 T, below $T^I/2$, the heat capacity decreases exponentially (figure 4 (b)), which might suggest the possible presence of gap opening [2]. Hence, the curve below $T^I/2$ is fitted with the following equation

$$C = \gamma_0 T + A \exp(-\Delta/T) \ldots (3)$$

where $\gamma_0$ is the electronic contribution in magnetic region, $A$ is a coefficient and $\Delta$ is the estimated gap opening value. The value of $\Delta$ obtained from fitting is ~ 14.3 ± 0.3 K (1.2 meV).

In order to further authenticate this gap opening, we tried to fit the DC susceptibility curve at 0.005 T obtained under ZFC condition. It is to be noted that this curve also decreases exponentially with temperature below $T^I/2$ (lower inset of figure 4 (b)). The curve is fitted with the following equation [2]

$$M/H\,(T) = M/H\,(0) + B\,\exp\,(-\Delta/T) \quad\ldots\,(4)$$

where $M/H\,(0)$ is zero temperature DC susceptibility and $B$ is a coefficient. The observed value of $\Delta$ is $\sim 15 \pm 0.3$ K, which is analogy with the heat capacity. Hence, it can be said that in this compound there is a presence of gap opening. This observation suggests that Fermi surface changes and it results in the evolution of new Brillouin zone boundaries.

**Resistivity and Magnetoresistance: Study of change in Fermi surface**

Temperature (1.8 – 50 K) dependent resistivity ($\rho$) behavior (in field range 0 - 14 T) of the compound is shown in figure 5 (a). With decreasing temperature, resistivity decreases due to the metallic behavior of compound, however, around $T^I$ it increases resulting in a maxima around $T^{II}$. Inset of figure 5 (a) shows the change in resistivity ($\rho_{max} - \rho_{min}$) as a function of magnetic field around $T^I$. It is observed that this change remains unaffected upto 3 T while it is suppressed above 6 T. This rise in resistivity indicates to the presence of gap opening due to changes in Fermi surface. Signature of such changes is also reflected in the behavior of the magnetoresistance (MR). Figure 5 (b) shows the magnetic field response of $\rho$ in temperature range of 1.8-15 K. With increasing magnetic field, the resistivity decreases due to localization of magnetic moment, as generally observed in Ce compounds [11]. Also, below $T^{II}$, the changes in resistivity is unaffected by the change in temperature. The $H^2$ dependence of MR (= $(\rho_H - \rho_0)/\rho_0$) at different temperatures is represented in figure 5 (c). It is noted that MR varies linearly with $H^2$ below 3 T. The $H^2$ behavior of MR is explained with equation [24, 25]:

$$-MR \sim \chi^2_{loc} H^2 \quad\ldots\,(5)$$

where $\chi_{loc}$ is the local magnetic susceptibility due to Brillouin contribution, where magnetization linearly varies with magnetic field. Above 3 T, MR deviates from $H^2$ due to non-linear variation of magnetization with magnetic field arising out from dominating exchange interactions of antiferromagnetic dipole moments. The slope of equation (5), $d(-MR)/dH^2$ is plotted as a function of temperature (inset of figure 5 (c)). From the figure, a peak is noted around $T^I$, and below this temperature $\chi_{loc}$ decreases. The variation in local moments is related with the changes in electronic properties of the compound [26]. Also, alteration of electronic properties can be a possible cause of the change in Fermi surface. Additionally, as stated before, below $T^{II}$ the change in MR is insignificant and this observation is further analyzed by Kohler's rule [27]. The Kohler's rule is derived from the semi-classical transport theory of Boltzmann equation. The scaling of Kohler's rule holds if there is only one species of charge carriers and scattering rate ($\tau$) is similar at all points of the Fermi surface. The field dependence resistivity is then included in the quantity $\omega\tau$, ($\omega$ is the frequency at which the magnetic field causes the charge to sweep across the Fermi surface). As the resistivity at zero field is directly dependent on $\tau$, the field dependence of MR with different $\tau$ due to different temperature can be related by rescaling the magnetic field by zero field resistivity [28]. Figure 5 (d) shows the Kohler plot in temperature range of 2-15 K. The equation for Kohler plot is as follows [29]:

$$\text{MR} = \Delta\rho/\rho(0) = f(\omega\tau) = f(H/\rho(0)) \ldots\ldots (6)$$

As noted from the figure above 5 K, the Kohler's rule is violated, while below 5 K the scaling shows a single curve (inset of figure 5 (d)). This observation suggests the reconstruction of Fermi surface (across $T^I$) which affects the symmetry of magnetic state. This leads to variation of electronic properties with temperature resulting in distinct scattering rate. However, this variation is insignificant at temperature less than $T^I/2$, indicating that the symmetry might be preserved, resulting in observation of scaling [28]. Hence it can be said that a new order parameter which develops around $T^I$, grows with decreasing temperature and is stabilized below $T^I/2$ because of some spin rearrangement. It results in preservation of symmetry of the magnetic state.

**Arrott plots: Analysis of order parameter through two-order-parameter model of Landau free energy theory**

Landau free energy density theory is interesting for investigation of the order parameter associated with higher order magnetization arising due to lowering of symmetry of magnetic state. The free energy density $F$ (in term of the two order parameters $M$ and $M_Q$) can be expressed in term of scalar quantity as [30, 31]:

$$F(M, M_Q) = c/2\ M^2 + d/4\ M^4 + l/2\ M_Q^2 + m/4\ M_Q^4 + n/2\ M^2 M_Q^2 - HM \ldots (7)$$

where $M_Q$ is the order parameter associated with quadrapole moment. $M_Q$ couples biquadratically with $M$. The coefficients $c$, $d$, $l$, $m$ and $n$ depend on temperature. $M$ couples linearly with the magnetic field. As in this compound the magnetic anisotropy plays a minimal role, it is expected that the scalar expression of Landau free energy theory will be valid. At zero field, the minima of $F$ corresponds to either paramagnetic state ($M = M_Q = 0$) or one of the possible following magnetic state: (i) $M = 0$, $M_Q \neq 0$; (ii) $M \neq 0$, $M_Q = 0$; (iii) $M \neq 0$, $M_Q \neq 0$. In this compound, magnetization studies suggest that for $H = 0$, possibility (i) might not occur because $M_Q$ is related with $M$. The conditions (ii) and (iii) are possible which is differentiated in the presence of external applied magnetic fields. Hence, for magnetic state, $M \neq 0$, $M_Q = 0$ for $H \neq 0$, the equation for minima of $F$ ($dF/dM = 0$) is described in form of Arrott plots as:

$$H/M = c + d\,M^2 \dots (8)$$

where $c$ and $d$ are the slope and intercept of equation respectively. For magnetic state, $M \neq 0$, $M_Q \neq 0$ for $H \neq 0$, the equation of Arrott plots for minimization of $F$ with respect to $M$ and $M_Q$ ($dF/dM = dF/dM_Q = 0$) is modified as:

$$H/M = c^* + d^*M^2 \dots (9)$$

where $c^*(T) = c - l\,n/m$ and $d^*(T) = d - n^2/m$ are slope and intercept of equation respectively. Figure 6 (a) and (b) shows the virgin curves of the isothermal magnetization at selected temperatures in the range 1.8 – 15 K. Figure 6 (c) and (d) shows the Arrott plots ($H/M$ versus $M^2$) extracted from these curves. In paramagnetic region, it is observed that the slope is positive which changes to an arc with negative slope ~ $T^I$. With decreasing temperature, the negative slope persist still 1.8 K. For obtaining the values of the slope and intercept, for both positive and negative slopes, we fit the equation (8) and (9) respectively. Figure 6 (e) and (f) shows the temperature dependent ($c$, $c^*$) and ($d$, $d^*$) respectively. The negative slope ($d^*$) increases with decreasing temperature. As reported in Ref [32], the change of slope from positive to negative is obtained for the first order magnetic transition and it is found that negative slope decreases with the decreasing temperature, which is in contrast to that observed for this compound. Such opposite behavior has been studied by taking into account the asymmetry of the free energy with respect to the magnetization [33]. Hence, it can be said that the presence of $M_Q$ lower the symmetry of the magnetic state which results in changeover of $c$ and $d$ to $c^*$ and $d^*$ respectively. This develops the new parameter which is second order in nature. The changeover from the positive to negative slope is an indication of development of a new second order parameter due to change of symmetry of the magnetic state. The change of symmetry occurs due to presence of higher order magnetization.

**Magnetic field – Temperature phase diagram of CeGe**

Figure 7 show the *H-T* phase diagram, which illustrates the possible presence of five phases in CeGe. The points in the figure are estimated from the ordering temperatures, obtained from the magnetization curves. In this compound, the phase I is present between $T^I$ and $T^{II}$, below 0.1 T. In this phase quadrapole moment competes with ordered antiferromagnetic dipolar moments. A new order parameter driven by quadrapole-quadrapole interactions develops which increases with decreasing temperature and lowers the symmetry of the magnetic state. It results in opening of a gap due to evolution of the Fermi surface. Phase II is below 0.1 T and $T^{II}$. In this phase, the interactions are dominated by quadrapole-quadrapole coupling which stabilizes the order parameter. The symmetry of the magnetic state is conserved in this phase. The observed ferromagnetic component arises due to the presence of short ranged correlations among quadrapolar magnetic moments. Phase III of the compound is between 0.1 to 1 T and below $T^I$. In this phase, dipole-dipole interactions dominate over the quadrapolar interactions. The region above 1 T and below $T^I$ is phase IV of the compound. In this phase, long ranged antiferromagnetic ordering is observed due to strong dipole–dipole interactions. The exchange energy of long range antiferromagnetic ordering is mediated via conduction electrons. Increase in magnetic field enhances the exchange interaction energy strength resulting in the observation of the long ranged ordering. The onset of this ordering suppresses the ferromagnetism in this phase. Phase V is the paramagnetic phase of the compound and it is above $T^I$.

**Summary**

In summary, our study on CeGe compound reveals that the presence of higher order magnetization in this compound, results in the development of a new order parameter. Presence of higher order magnetization indicates the lowering of symmetry of the magnetic state and opens a partial gap due to changes in Fermi surface. The evolution of Fermi surface is also substantiated from the $H^2$ dependent magnetoresistance behavior and Kohler's plots. The observed order parameter (which is second order in nature) is confirmed from the Landau free energy theory. Further, for an unambiguous identification of the new order parameter microscopic probes like resonant x-ray scattering, field dependent neutron diffraction, NMR etc. are warranted.

**Methods**

High quality polycrystalline compound CeGe is prepared by arc melting the stoichiometric amounts of high purity (>99.9%) Ce and Ge in an atmosphere of argon. The characterization of compound is carried out by x-ray diffraction (Cu K$α$) pattern obtained using Rigaku Smart Lab instruments. Temperature dependent magnetization measurements are performed using Magnetic Property Measurement System (MPMS), while, temperature and magnetic field dependent heat capacity and resistivity measurements are performed using Physical Property Measurement System (PPMS), both from Quantum design, USA. Magnetization, heat capacity and resistivity measurements are carried out on pellets of specific shapes.

**Acknowledgements**

The authors acknowledge the experimental facilities of Advanced Material Research Centre, IIT Mandi. Financial support from IIT Mandi is also acknowledged.

**Figures**

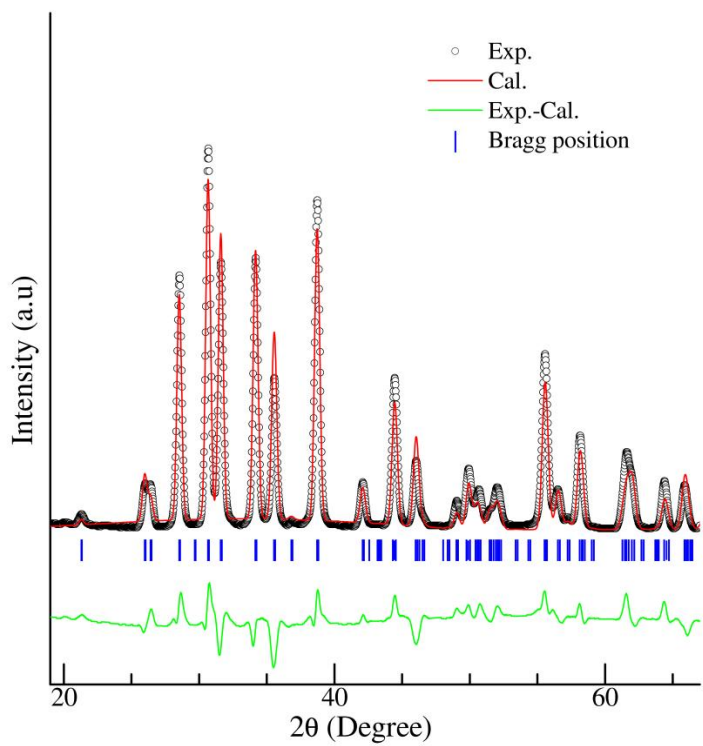

**Figure 1:** Rietveld refined powder x-ray diffraction pattern of CeGe. The difference between the observed and the experimental pattern and the Bragg position are also shown.

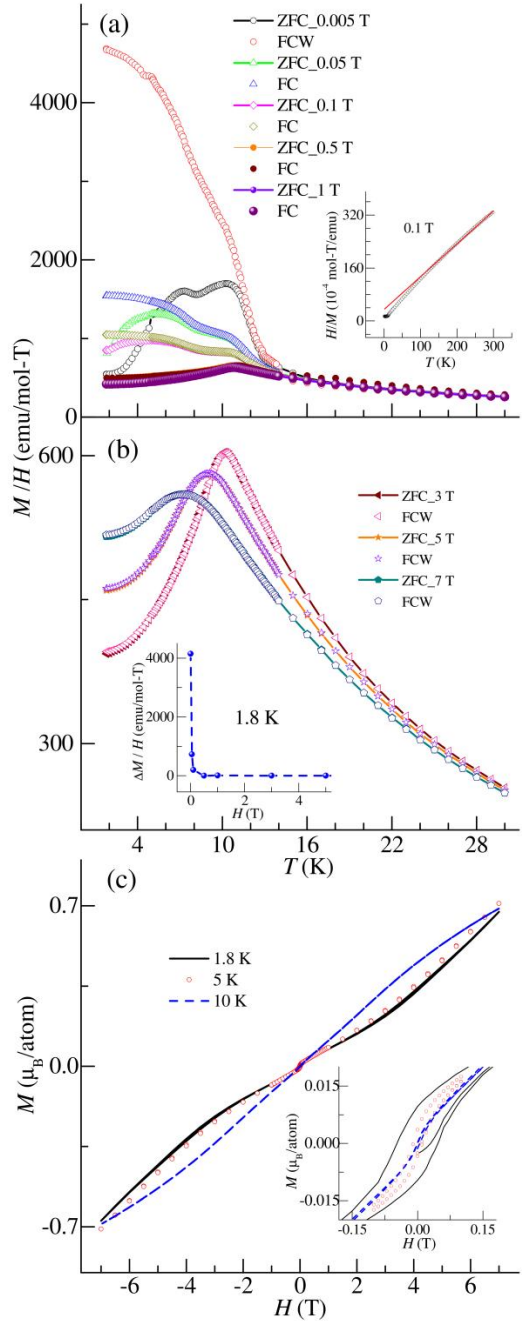

**Figure 2:** (a) and (b) Temperature (*T*) dependent DC susceptibility (*M*/*H*) in field range of 0.005 - 7 T under ZFC and FCW condition. Inset (a): Temperature dependent *H*/*M* at 0.1 T. Red curve shows the Curie Weiss law fitting. Inset (b): Magnetic field (*H*) dependent Δ*M*/*H* (Δ*M* = $M_{FC}$ − $M_{ZFC}$) at 1.8 K. (c) Isothermal magnetization (*M*) curves as a function of magnetic field (7 to -7 T) at 1.8, 5 and 10 K. Inset: Expanded form of the same curves at low fields.

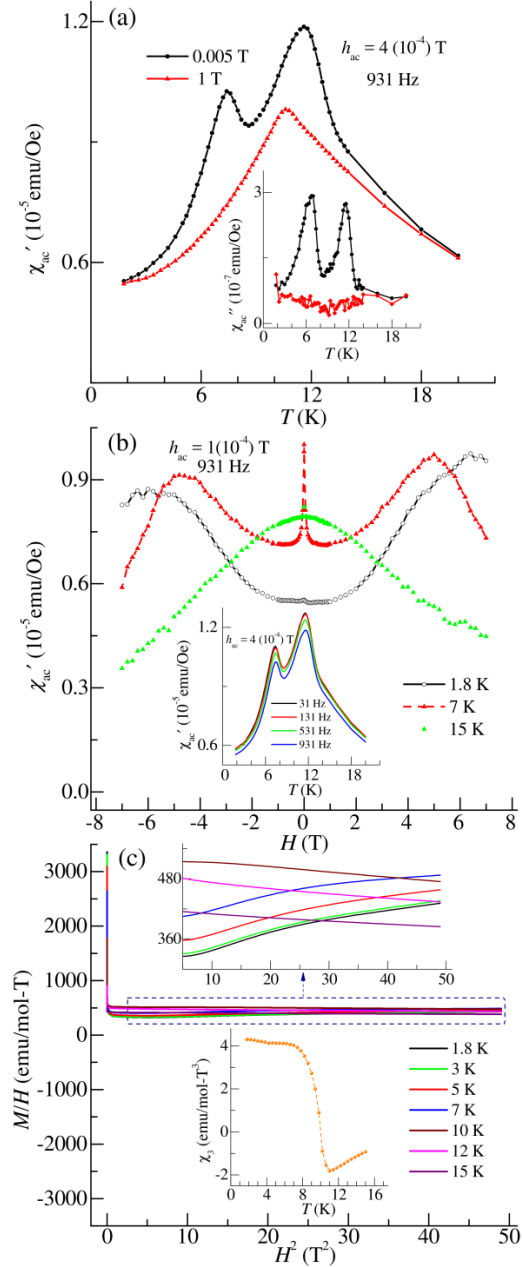

**Figure 3:** (a) Temperature dependent real part of AC susceptibility ($\chi'_{ac}$) at 4 ($10^{-4}$) T AC field and superimposed DC field of 0.005 and 1 T; at 931 Hz. Inset: imaginary part of AC susceptibility ($\chi''_{ac}$) under similar condition. (b) DC magnetic field (7 to -7 T) dependent $\chi'_{ac}$ at 1.8, 7 and 15 K and 931 Hz in presence of 1 ($10^{-4}$) T AC field. Inset: Temperature response $\chi'_{ac}$ in different frequencies in presence of 4 ($10^{-4}$) T AC field and superimposed DC field of 0.005 T. (c) $H^2$ dependent $M/H$ at different temperatures. Upper inset: Expanded portion of the same curves in field range 2-7 T. Lower inset: Temperature response of $\chi_3$.

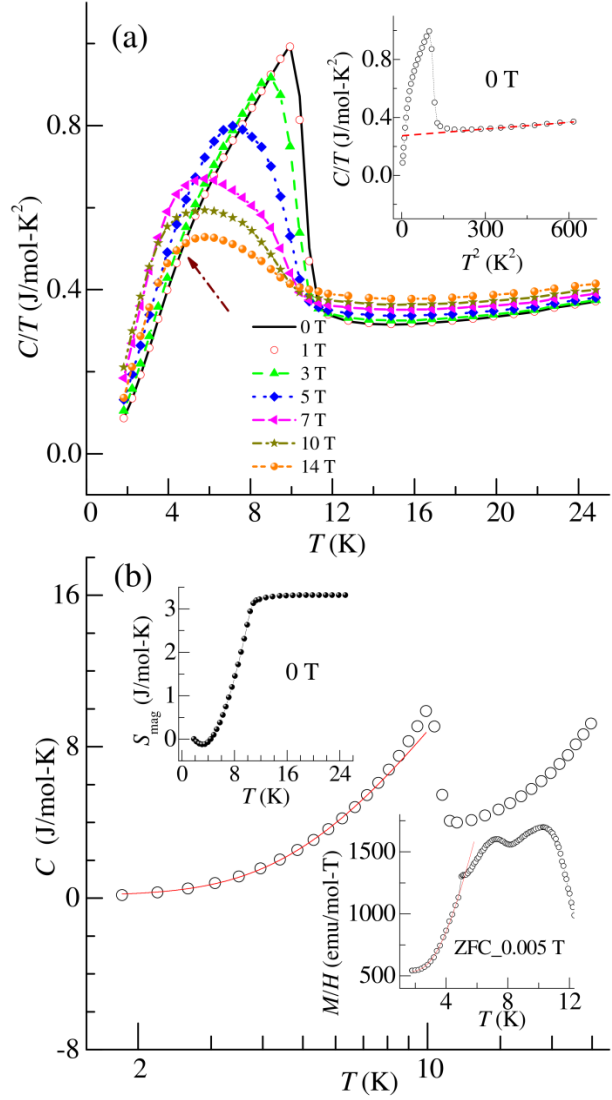

**Figure 4:** (a) Heat capacity divided by temperature (*C*/*T*) plotted as a function of temperature in field range of 0-14 T. Arrow shows the deviation in *C*/*T* near 7.6 K at 0 T. Inset: $T^2$ dependent *C*/*T* at 0 T. Red curve shows the linear fitting above the ordering temperature, which is extrapolated to the lowest temperature. (b) Temperature (logarithmic scale) dependent heat capacity at 0 T. Red curve shows the curve fitting of equation (3). Upper inset: Temperature dependent magnetic entropy ($S_{mag}$) at 0 T. Lower inset: Temperature dependent *M*/*H* at 0.005 T (ZFC mode). Red curve shows the fitting of equation (4).

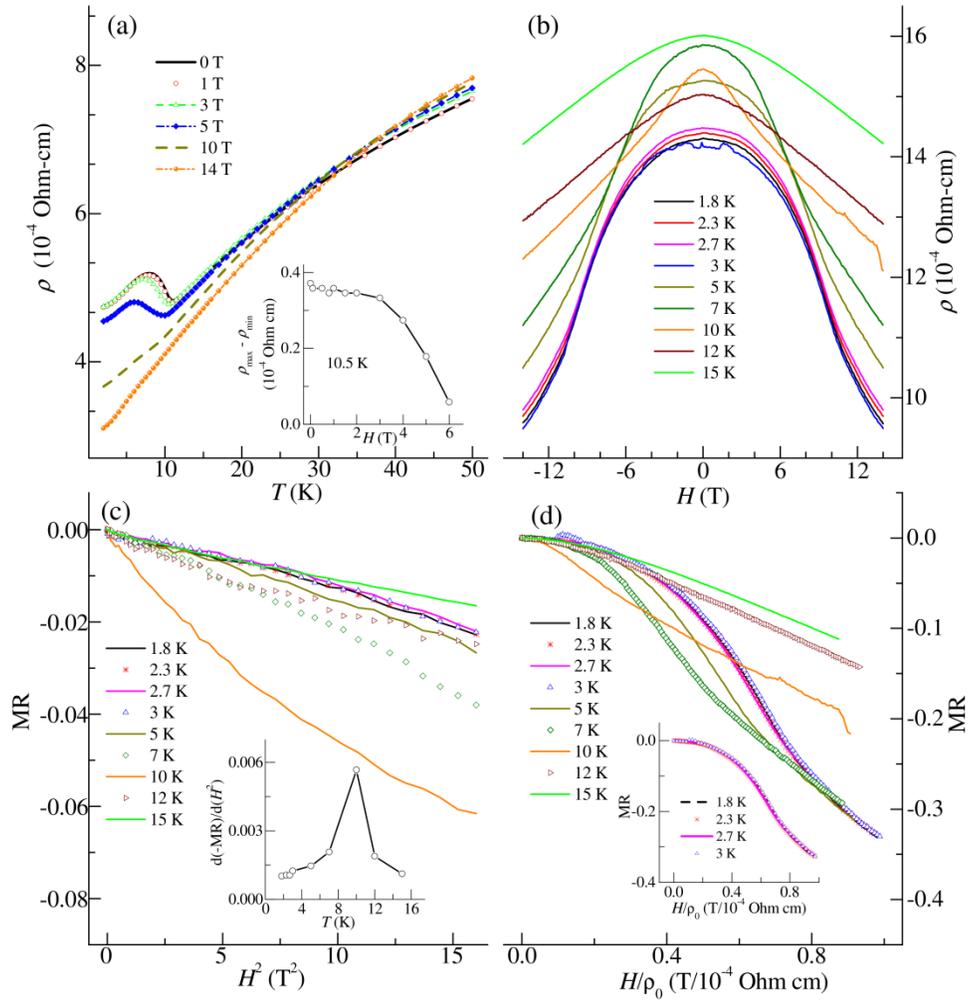

**Figure 5:** (a) Temperature dependent resistivity (ρ) in field range of 0-14 T. Inset: Magnetic field dependent resistivity ($\rho_{max} - \rho_{min}$) at $T^I$. (b) DC magnetic field (14 to -14 T) dependent resistivity in temperature range of 1.8 – 15 K. (c) $H^2$ dependent magnetoresistance (MR) at similar temperatures. Inset: Temperature dependent d(-MR)/d($H^2$). (d) $H/\rho_0$ dependent MR at similar temperature. Inset: Scaling of Kohler's rule.

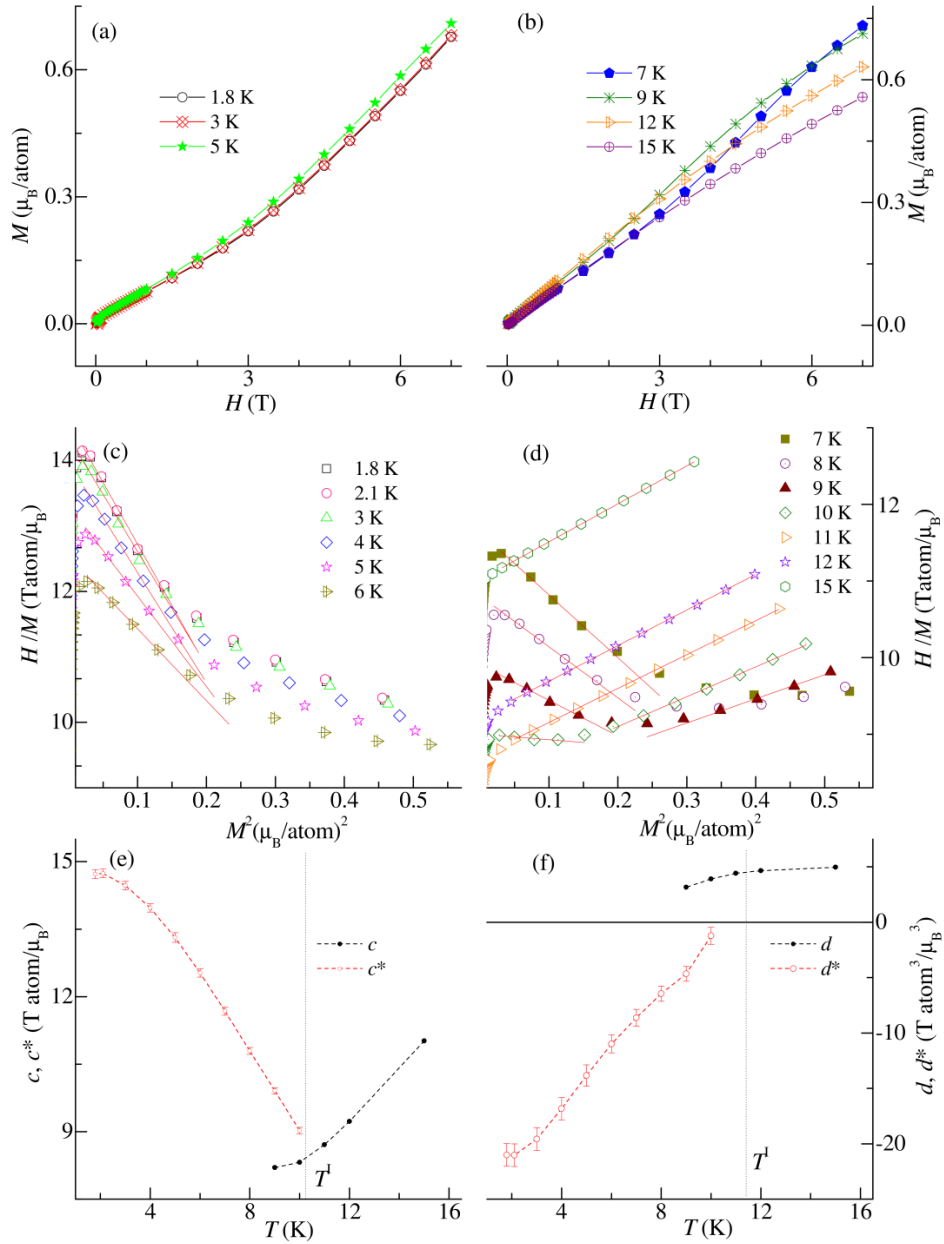

**Figure 6:** (a) and (b) Virgin isothermal magnetization (*M*) curves as a function of magnetic field (0 to 7 T) at different temperatures. (c) and (d) *H*/*M* versus $M^2$ in temperature range of 1.8 – 15 K. Red curve shows the linear fitting of the plots. (e) and (f) Intercepts (*c*, *c**) and slopes (*d*, *d**) plotted as a function of temperature.

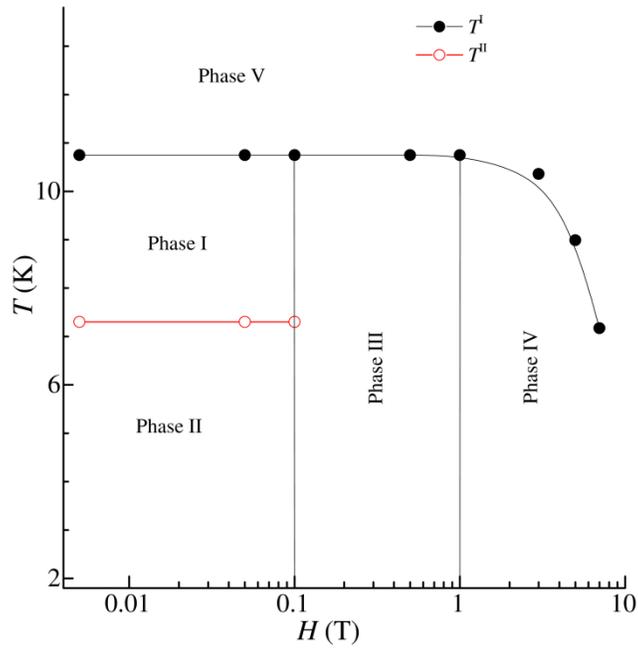

**Figure 7:** Temperature-Magnetic field phase diagram of CeGe.

.